\documentclass[conference]{IEEEtran}
\IEEEoverridecommandlockouts
\usepackage{cite}
\usepackage{amsmath,amssymb,amsfonts}
\usepackage{algorithmic}
\usepackage{graphicx}
\usepackage{textcomp}
\usepackage{url}
\usepackage{xcolor}
\def\BibTeX{{\rm B\kern-.05em{\sc i\kern-.025em b}\kern-.08em
    T\kern-.1667em\lower.7ex\hbox{E}\kern-.125emX}}
\begin{document}

\title{Partially Trusting the Service Mesh Control Plane}
\author{\IEEEauthorblockN{\rm Constantin Adam\\ Abdulhamid Adebayo\\ Hubertus Franke\\ Edward Snible\\ Tobin Feldman-Fitzthum\\ James Cadden}\\
IBM T.J. Watson Research Center
\and
{\rm Nerla Jean-Louis}\\
UIUC
}

\maketitle

\begin{abstract}
Zero Trust is a novel cybersecurity model that focuses on continually evaluating trust to prevent the initiation and horizontal spreading of attacks. A cloud-native Service Mesh is an example of Zero Trust Architecture that can filter out external threats. However, the Service Mesh does not shield the Application Owner from internal threats, such as a rogue administrator of the cluster where their application is deployed. In this work, we are enhancing the Service Mesh to allow the definition and reinforcement of a Verifiable Configuration that is defined and signed off by the Application Owner. Backed by automated digital signing  solutions and confidential computing technologies, the Verifiable Configuration allows changing the trust model of the Service Mesh, from the data plane fully trusting the control plane to partially trusting it.  This lets the application benefit from all the functions provided by the Service Mesh (resource discovery, traffic management, mutual authentication, access control, observability), while ensuring that the Cluster Administrator cannot change the state of the application in a way that was not intended by the Application Owner.
\end{abstract}


\section{Introduction}
\label{sec:intro}

With the widespread adoption of hybrid cloud deployments, data and application confidentiality remain of paramount concern, especially in regulated industries, such as finance, telecommunications, or healthcare.  Zero Trust and Confidential Computing are recent initiatives aiming to address this problem.  Zero Trust is a new cybersecurity model that covers a collection of concepts and ideas designed to minimize uncertainty in enforcing accurate, least privilege, per-request access decisions in information systems and services in the face of a network viewed as compromised.

Confidential Computing delivers enclaves or trusted execution environments (TEE), consisting of either a portion of an application or an entire virtual machine. Confidential Computing enables users to compute without exposing their application or data to the operator of the underlying infrastructure. They do so by creating an orthogonal privileged domain separate from the one controlled by the operating system or hypervisor. This privileged domain is often rooted in hardware and firmware; it can be brought up on a system and controlled and attested independently of the host via trusted hardware (the Platform Security Processor for AMD, the Management Engine for Intel, the Ultravisors in IBM Z, and P). In a cloud environment, the tenant may access and attest the TEE independently of the state of the host system or other tenants, thus considerably reducing the trust footprint for running it.

Confidential Computing thus provides an ideal environment for reducing trust to the levels required by a Zero Trust environment by providing at rest, in motion, and use protection for data inside a security enclave. This ideal security model, which consists of operation attestation at the beginning and booting of encrypted images, gives a highly confidential guarantee that by the time the container is running, the cloud provider has no access to any data running in that container.

However, deployment in the real world introduces a new degree of complexity. Most of today's hybrid cloud-native applications are distributed and often count on third-party tools, like sidecars and service meshes, to implement various security or traffic management functions. For these additional functions to work, the Confidential Computing security model of the application has to be relaxed to partially trust the third party control plane and allow it to interact with the application containers located inside the security enclave.

The main contribution of this paper is to formally define this concept of Partial Trust - how Partial Trust can be created, established, identified, and enforced.  We describe the procedure of identifying the line between trusted and untrusted information using Istio, a popular Service Mesh implementation, as the motivating example.  The concept of Partial Trust departs from traditional orchestration and Service Mesh architectures, where the control plane is fully trusted, and the API between the control and the data plane can be used to create new containers, configure the network, or get logging information.


The rest of the paper is structured as follows.  Section~\ref{sec: bg-threat} provides a brief overview of the Service Mesh, focusing on Istio, and it also defines the threat model that we are addressing in this work.  Section~\ref{sec:rel-work} discusses in more detail the supporting technologies for our solution, and reviews related work.  Section~\ref{sec:overview}  provides a system overview.  We discuss how this approach can help defending against various attack techniques in Section~\ref{sec:discussion}, and conclude in Section~\ref{sec:conclusion}.

\section{Problem}
\label{sec: bg-threat}

Zero Trust Architecture (ZTA) is an Enterprise CyberSecurity Architecture based on Zero Trust principles. Extensive guidance on transitioning to and implementing ZTA is covered in \cite{nist-zta}.  The Istio Service Mesh is a fundamental component in implementing ZTA for cloud-native applications. The Service Mesh provides a multitude of functions, such as mutual authentication and encrypted communication between services, policy propagation and enforcement, certificate rotation, that allow to filter out external attackers and threats.

\begin{figure*}[h]
\centering
\includegraphics[width=\linewidth]{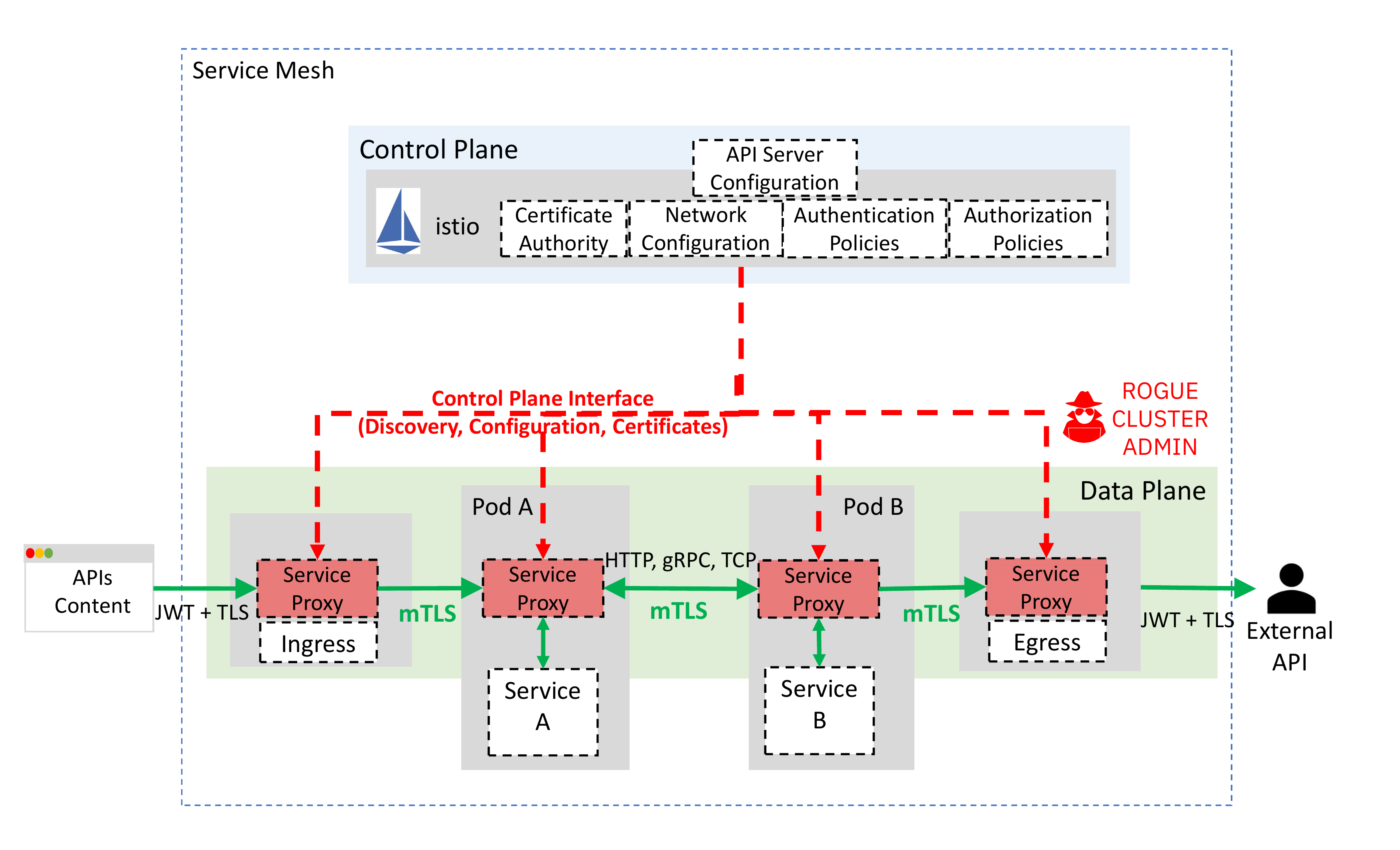}
\caption{Internal Service Mesh Attack}
\label{fig:rogue-cluster-admin}
\end{figure*}

However, the Service Mesh brings new security risks into the system, as it does not provide adequate protection for an Application Owner against the Cloud or Kubernetes administrators.  Running each application pod in a security enclave shields the pods from attacks originating from the host Operating System. However, as shown in Figure~\ref{fig:rogue-cluster-admin}, insiders can use the interface between the control and data planes of the Service Mesh to escalate their privileges and abuse them by getting insights into the current state of the Service Mesh, and then misconfiguring it, to change the behavior of the connections, and of the traffic that is routed along those connections.  And even if the insiders do not act maliciously, many regulations prohibit the administrators and the operations personnel from having access to the applications and their data. 

In the rest of this section, we highlight the fundamentals of the Service Mesh architecture with Istio and discuss the threat model.

\subsection{Service Mesh with Istio}
Service Mesh provides a dedicated infrastructure for supporting services in microservices-based applications. Cloud-based and enterprise applications adopt the microservices architecture because of its agility, scalability and increasing availability of automation tools \cite{chandramouli2020building}. Service Mesh is designed to standardize the run-time operations of an application such as service discovery, load balancing,  traffic management \cite{khan2017key}, interoperability \cite{truong2018service} and dependency control \cite{esparrachiari2018tracking}. Istio \cite{istio}, started as a collaboration between IBM, Google and Lyft, Linkerd \cite{linkerd} a 100\% Apache-licensed open source product, and Consul Connect \cite{consul},  a Service Mesh from Hashicorp, are the most popular service meshes in the market today.

The increased adoption of the microservices architecture on a distributed system like Kubernetes results in increased management complexities. Concepts such as service discovery, routing, fail-over become prominent as distributed services communicate with each other. Istio provides an efficient mechanism to secure and monitor these microservices through an array of Envoy proxies. Envoy is an OSI L3 and L4 network proxy that uses a chain of network filters to perform connection handling. Envoy also supports L7 layer filters for HTTP traffic. Many of Istio's features are enabled by the underlying built-in attributes of the Envoy proxies.

Istio consists of two components regardless of its deployment pattern: the data plane and the control plane. The data plane is the network of microservices whose communication is routed through Envoy proxies that run alongside the services. These proxies create the mesh network for the services. Envoy allows the application of HTTP, gRPC, WebSocket, and TCP routing rules for fine-grained traffic control. Envoy also supports security policies and access control for services.

The management of the proxies is done from the control plane. 
\emph{Istiod} dynamically configures the Envoy sidecar proxies using a set of discovery APIs, collectively known as the xDS APIs.  The control plane provides configuration, discovery, and certificate management for the Envoy proxies at runtime.  \emph{Istiod} is the component that converts routing rules specified in high-level formats such as YAML into Envoy-specific configurations and propagates them to the sidecars. As part of \emph{Istiod}, \emph{Pilot} is responsible for service discovery, \emph{Galley} for configuration, and \emph{Citadel} for certificate generation.  A fourth component, \emph{Mixer}, providing extensibility has been deprecated. The functionality provided by Mixer is now offered through WebAssembly, a sandboxing technology which can be used to extend the Istio proxy (Envoy). The Proxy-Wasm sandbox API replaces Mixer as the primary extension mechanism in Istio.  

\emph{Istiod} provides end-user and service-to-service authentication with its built-in identity and credential management component. Istio offers peer authentication and request authentication. For peer authentication, Istio offers mutual TLS for service-to-service authentication. Istio uses JSON Web Token (JWT) validation for end-user authentication. Mesh, namespace, and service-level access control to the Envoy proxy inbound traffic is enforced using authorization policies.

\subsection{The Threat Model}
The adversary in our model is the Cluster Administrator.  Because the adversary has the administrative rights to the cluster, they can create new objects, and delete or modify the existing objects in the Kubernetes cluster.  We assume that the workload certificates are provisioned using a custom certificate authority, not the Istio certificate authority, and the Envoy proxies are configured to only accept certificates issued by the custom certificate authority.  The adversary cannot, therefore, tamper with the certificate management.  Also, because the application pods run in security enclaves controlled by the Application Owners, the adversary cannot see or modify the contents of the application pods.

The adversary can interact with the application pods in three ways. First, they can create, modify, or delete CRDs in the cluster. The Istio daemon will be notified of actions and, as a result, will send the service proxies messages (over their xDS API interface) that will trigger the reconfiguration of the service proxies. Second, the adversary can create new pods inside the cluster, label them as application components, and establish communication with the application pods through these falsely advertised services. Third, the adversary can poison the Domain Name System (CoreDNS) or the ARP cache of the cluster. 

We cover the first scenario described above, when the adversary creates, deletes, or modifies CRDs. For this purpose, we are using Cosign, a digital signing and verifying technology. We are not covering man-in-the-middle attacks here, where the adversary can store the signatures associated with different policies and reply later to create configuration resources without the Application Owner's approval. Since communication in the Service Mesh is encrypted based on keys and certificates, a Custom Certificate Authority can address the second and the third scenarios. Introducing a custom certificate authority, that resides in the control plane that stores and distributes the keys and certificates used in the data plane, prevents communication in the Service Mesh from being exposed to malicious traffic and manipulations \cite{el2019guiding}. However, these scenarios are beyond the scope of this paper.

\section{Related Work}
\label{sec:rel-work}
MarbleRun~\cite{marblerun} is a state of the art security framework for service meshes.  MarbleRun guarantees that the topology of a distributed application adheres to a manifest specified using JSON. MarbleRun verifies the integrity of services, bootstraps them, and sets up encrypted connections between them. If a node fails, MarbleRun will seamlessly substitute it with respect to the rules defined in the manifest.  The main difference between MarbleRun and our approach is that the control plane in MarbleRun runs in a security enclave, and, as such, its administration and management fall under the responsibility of the Application Owner.  SCONE~\cite{199364} is another notable secure container environment.  SCONE compiles unmodified source code into an enclave application binary using an SGX-aware musl-libc and/or run unmodified Alpine Linux binary.  SCONE provides comprehensive encryption protection not only for files, but also for environment variables and input parameters. SPIFFE~\cite{spiffe} (and its implementation, SPIRE)  solves the problem of assigning identities to workflows.  SPIFFE allows to name each service, or process and gives them an identity to communicate with each other that is independent of the underlying stack (virtual machine, platform, etc.).  SPIFFE performs a function that is complementary to our approach, as it does not attempt to control the policies that are distributed to the data plane of the service mesh. 

Current technologies for confidential container clouds (e.g. encrypted containers) are limited to security threat models that require users to trust cloud operators who provide the host operating system hypervisor and the Kubernetes control plane \cite{Lum-encrypted-images}. Technologies to prevent user-to-user attacks are based on wrapping an encrypted containers inside a virtual machine (e.g. Kata containers \cite{kata-containers}). This improves the container security by increasing isolation between different users on the same compute nodes, but does not guarantee the confidentiality of the compute and data container content with respect to the hypervisor and to the control plane.

Confidential computing hardware, such as AMD SEV (~\cite{amd-sev}), Intel TDX (~\cite{intel-tdx}) and IBM Z secure~(~\cite{ibm-z-secure}), aims at preventing malicious hypervisors to access cloud user data, by wrapping applications within a Virtual Machine (VM) using encrypted memory. These technologies provide different guarantees on isolation level, cache and memory access protection, and protection against snooping and hardware-based attacks \cite{hardware-sol}. Several ongoing projects leverage this hardware to build and deploy confidential and secure containers based on hardware-supported VM encryption and full-stack confidentiality and security. 

Technologies like Intel SGX and TEEs are used for a variety of applications. In \cite{201548} a TEE is used to protect sensitive TOR operations and prevent attackers who own relays from getting access to sensitive data. In \cite{DBLP:journals/corr/abs-1902-04413} a limited TensorFlow API is implemented in an enclave to reduce the security risks of training machine learning models using sensitive data in the cloud. In \cite{8418608}, the authors present an efficient database implementation within an enclave by decoupling compilation of queries outside the enclave and execution of queries within the enclave. In \cite{seng} the authors are able to create a firewall that can attribute malicious behavior to applications in the system while maintaining the privacy of the applications using SGX.

In \cite{mesh-security-1} the authors examine the Service Mesh model's security in different scenarios with varying skill levels of service administrators and attackers. They show the risks for disruption, manipulation, and takeover by attackers and that mitigating these adversarial goals requires a high level of skill and manual configuration by a service administrator. Even in the best case scenario there are still risks to the system. Our design mitigates these risks by using hardware backed secure enclaves for greater isolation and allowing clients to have power to override any errors made by a system administrator. 

In \cite{mesh-security-2} the authors analyze internet artifacts to understand the security practices for Kubernetes better. The authors acknowledge the limitations in the study's sample size, give a non-exhaustive list of guidelines for administrators, and emphasize the great responsibility that administrators bear in collecting and applying this information. In \cite{mesh-security-3} the authors test the performance trade-offs for adding similar hardware isolation to Kubernetes run-times. While the system takes a performance hit to add this extra isolation, we believe that the increased security guarantees justify this cost. 

Although Intel SGX has seen widespread use in literature, it and similar technologies have been shown to have many vulnerabilities and published attacks \cite{sgxattacks}.  It has also been shown to be susceptible to side channel attacks that allow untrusted parties to infer secrets based on cache usage and other channels \cite{sidechannel}.

This work builds upon the capability of providing security enclaves where the application containers and the sidecar containers can bootup without exposing the contents of the image or the key used for the image decryption to the Cluster Administrator. 

Digital signatures are often used to provide integrity in cloud-based systems. A component such as an image or a configuration can be verified digitally as originating from an organization using a signed certificate. In \cite{kudo2021integrity}, the signature of a Kubernetes resource is verified at the admission controller to protect its integrity. The verification of the resource is based on the initial Kubernetes resource before it is internally modified. Change monitoring and change anomaly detection by identifying mutations based on rules was proposed in \cite{jin2010guest, zlatkovski2018new}.

Trust is an important consideration when transferring data between networked systems. It is critical to ensure the integrity of such data as it is moved from one ownership domain to another. More recently, container image signing has become a well sought after problem in the security landscape with the development of projects such as Docker Content Trust (DCT) \cite{dct} and Cosign \cite{cosign}. DCT enables a one-time check by the Docker client to verify that images pulled are signed on a Docker Notary server. In its original use case, Cosign was developed for raw image signing.  The Cosign project is part of the Sigstore \cite{sigstore} initiative. Sigstore is a set of community-driven tools used for digital signing, verification, and provenance checks when using open-source software. With Cosign, an image built in a CI pipeline can be signed, and anyone downloading the image would be able to verify that the image they are downloading was built by someone with access to the private key.

This paper proposes an approach to prevent unauthenticated configuration changes to the Istio Service Mesh by using signatures. Application Owner defines a list of configurable custom resources within the Service Mesh that should only be implemented if it passes the integrity check. The signature verification is performed at the envoy proxy after the initial CRD has been converted into its gRPC equivalence. This allows a variety of integrity checks to be performed for different business cases in the same Service Mesh.

\section{System Overview}
\label{sec:overview}

\begin{figure*}[h]
\centering
\includegraphics[width=\linewidth]{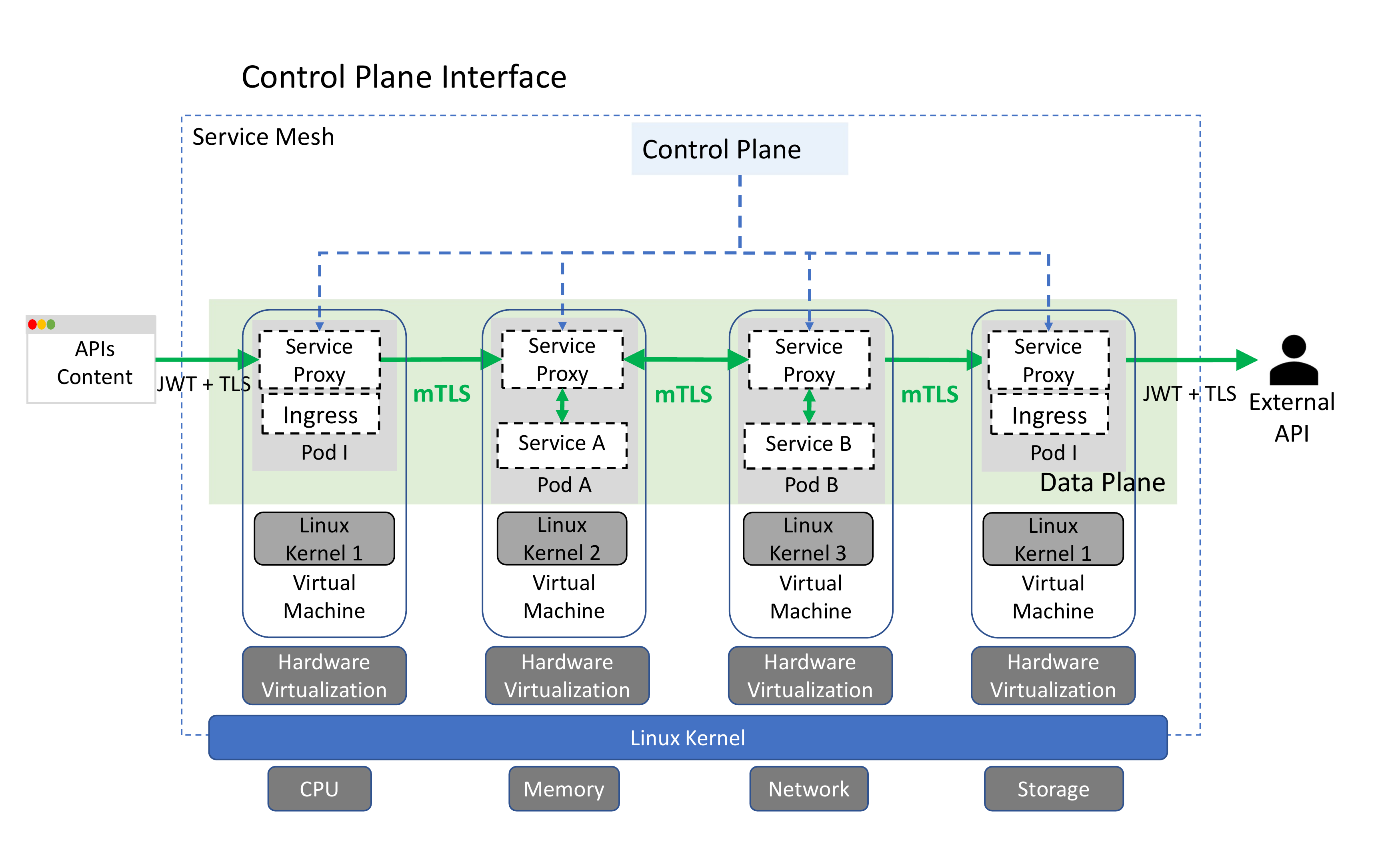}
\caption{Zero-Trust Architecture Stack}
\label{fig:zta-stack}
\end{figure*}

Figure~\ref{fig:zta-stack} shows the full deployment stack of our ZTA. Once attested hardware and a hardened runtime using secure containers (e.g., Kata containers) are in place, we focus on the Service Mesh and how to expand the individual security enclaves. Application micro-services and their sidecar containers run in a TEE that is under the control of the Application Owner and cannot be modified or tampered with by a Cluster Administrator. We look at a security model from the application owner's perspective, where the Cluster Administrator is the adversary, who can create new objects and delete or change the configuration of existing objects in Kubernetes, including CRDs that can change the behavior of the Service Mesh. To do so, we need to implement the following functionality:
\begin{itemize}
    \item Ensure that the system boots up securely, with a well-defined configuration,
    \item Allow the Application Owner to sign off (offline) on a specific set of policies, 
    \item Discard (in the sidecar containers that are attached to the Service Mesh) the configuration settings that have not been approved by the Application Owner (and cannot be verified using the Application Owner's security key)
\end{itemize}

In the rest of this section, we describe the implementation of each functionality listed above in detail.
\subsection{Secure Bootup}
\label{subsec:bootup}

Kata Containers facilitate the deployment of Kubernetes pods inside virtual machines. Confidential Computing technologies such as AMD SEV, Intel TDX, or IBM Z Secure Execution allow for creating encrypted VMs. Kata can be used with these VM enclaves to deploy a Kubernetes pod inside a TEE. With Kata, an entire pod can be deployed in an enclave, including sidecars such as the Istio Envoy Proxy. Some older Confidential Computing technologies, such as Intel SGX, have been used to deploy single containers inside of process-based enclaves\cite{SCONE} \cite{XContainers}. That said, Kata is not suitable for Confidential Computing without modification.

Without Confidential Computing, Kata Containers isolates one guest from another, but it is not designed to protect guests from a malicious cloud service provider or orchestrator. In Kata, information and commands cross the VM boundary from the host to the guest. For example, container images are downloaded on the host and then exposed to the guest via filesystem passthrough. Furthermore, the Kata Agent inside the guest operates on behalf of the orchestrator to start containers. The Kata Agent uses a VSock to communicate with the Kata Shim running on the host. The $Shim$ can issue commands to the Agent. Beyond starting containers, the Agent can also execute arbitrary commands inside the guest on behalf of the $Shim$. 

In Confidential Computing, the guest does not trust the host. The existing Kata architecture is not suitable for this. Thus,  a project called Confidential Containers (CC) extends Kata for use with VM Enclaves \cite{confidential-containers}. CC builds on Kata to support pulling encrypted container images from inside the guest to avoid exposing confidential information to the host. CC also introduces a configurable Kata Agent API so that the Agent can refuse specific commands from the host. CC provides an attestation flow for verifying the guest environment before provisioning sensitive workloads.

Kata Containers, and CC by extension, isolate a Kubernetes pod inside the VM. Each VM has its firmware and operating system isolated from other VMs containing other pods. Kata uses a minimal guest operating system. The VM boots to an $initrd$, pre-loaded with the Kata Agent. With CC, the guest must be measured before secrets can be provided. This entails the measurement of the guest firmware, $initrd$, kernel, and kernel parameters. The underlying Confidential Computing hardware must perform this measurement so that the measurement is linked to the hardware root of trust. Typically AMD SEV only measures guest firmware, but recent enhancements extend the guest firmware to measure further components \cite{sev-spec} \cite{kvm-forum}. By verifying the measurement of the guest, the guest owner can trust the Kata Agent. Memory encryption protects the Kata Agent at runtime. The Agent can then download, verify, and decrypt standard signed and/or encrypted container images. This flow allows for the secure provisioning of sensitive workloads and sidecars.

\subsection{Partially Trusted Control Plane}
\label{subsec:partial-trust}

\begin{figure*}[h!]
\centering
\includegraphics[width=\linewidth]{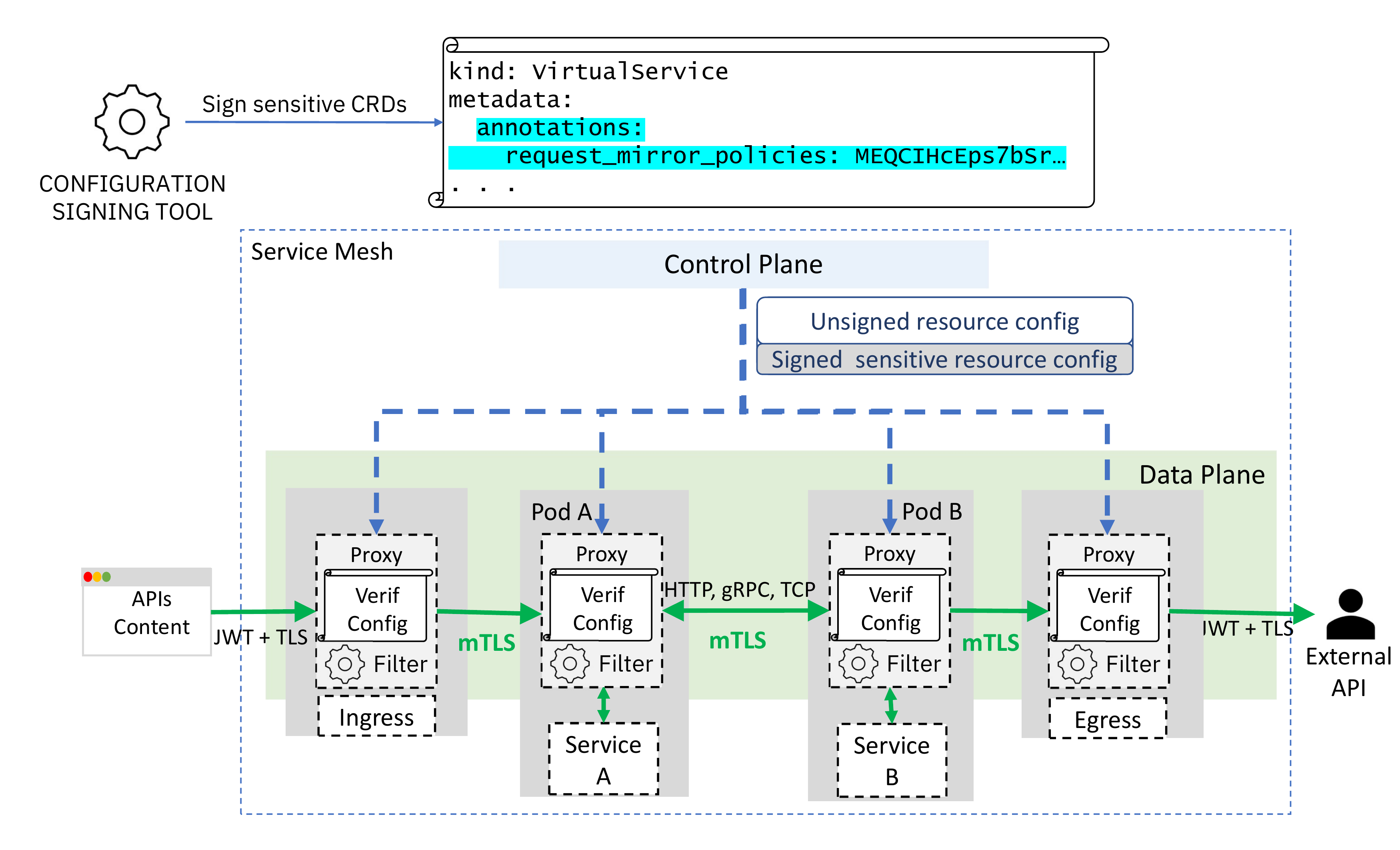}
\caption{Partial trust between data plane and control plane in the Service Mesh}
\label{fig:partial-trust}
\end{figure*}

Figure~\ref{fig:partial-trust} illustrates the partial trust between the data and the control planes in the Service Mesh.  

In the current Service Mesh implementations, the service located in the data plane trusts the resource configurations received from the control plane and modifies their state accordingly.  Our partial trust model splits the interface between the data and the control planes into a confidential part that requires further review of the configuration settings received from the control plane before processing them and a non-confidential API where the data plane immediately processes any resource configurations received from the control plane and updates its state.  

The API is partitioned through a Verifiable Configuration - a list of configuration settings requiring further review before processing the proxies.  Each proxy receives a Verifiable Configuration during bootup.  Upon receiving from the control plane a new message, a filter in the proxy examines the list of resources included in the message.  If any resources contain Verifiable Configuration policies, their content is checked using a signature and a security key or rejected.  Only the verified resource configurations  are forwarded to the proxy and processed by the proxy. 

\begin{figure*}[h]
\centering
\includegraphics[width=\linewidth]{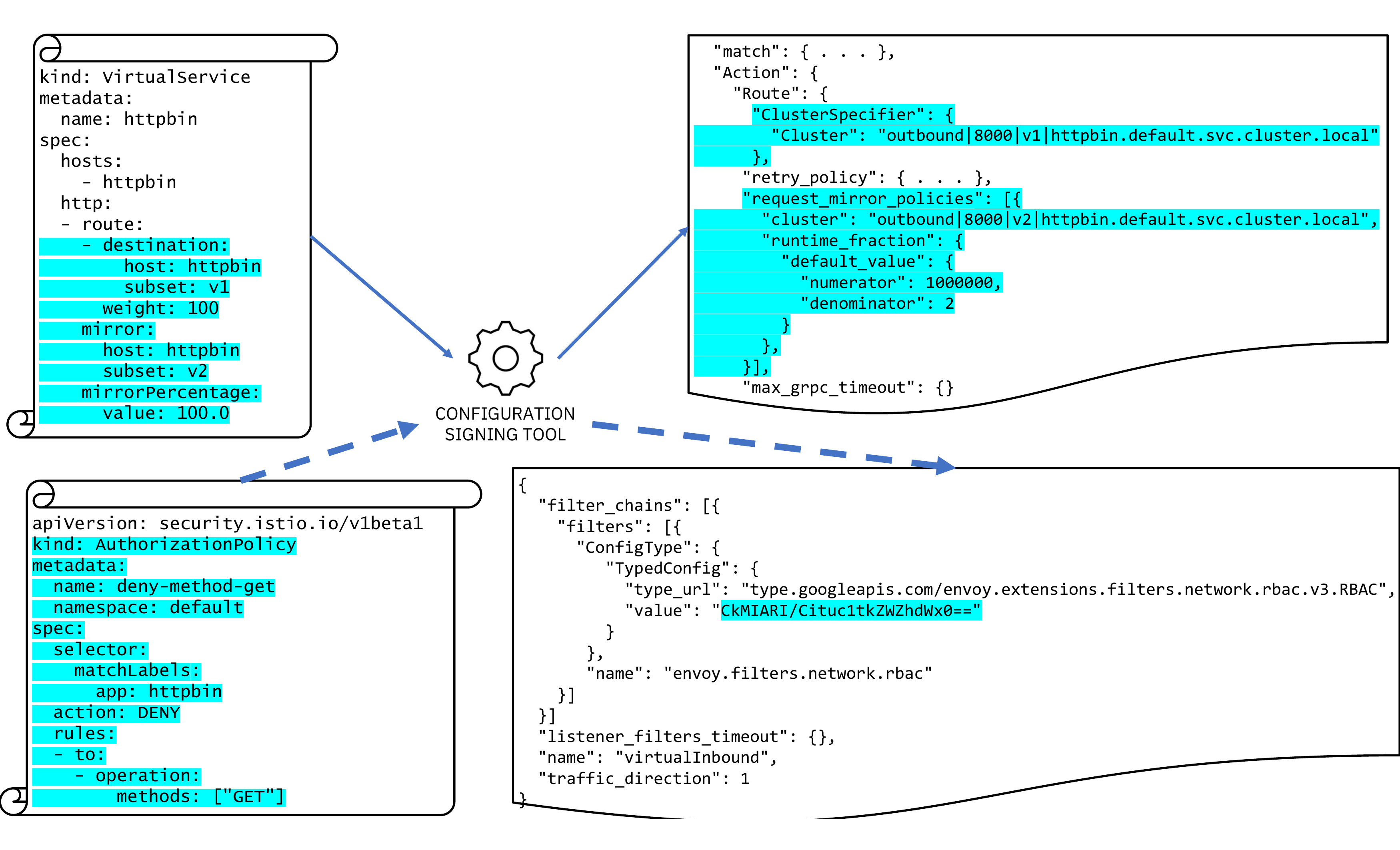}
\caption{Configuration Signing Tool}
\label{fig:config-sign-tool}
\end{figure*}

\subsection{Verifiable Configuration}
\subsubsection{Verifiable Configuration Definition}
\label{subsubsec:verifiable-config}
The Verifiable Configuration is defined as the set of all the configuration settings that must be signed and verified to be applied to the sidecar containers. For example, suppose the Verifiable Configuration includes traffic mirroring policies. In that case, all the fields of a route configuration resource that define a mirroring policy must be signed and verified before being applied. The Verifiable Configuration contains an enumeration of the policies that must be verified, e.g., it can include the request mirroring or authorization policies. However, it does not provide any specific conditions on the contents of those policies. In this way, the Application Owner can define any policies, with any conditions, for any target endpoints, and those policies will be approved if they can be verified using the public key of the Application Owner and signature of the policy, generated using Cosign.

\subsubsection{Signing a Configuration Change}
\label{subsubsec:sign-config}
\begin{figure*}[h!]
\centering
\includegraphics[width=\linewidth]{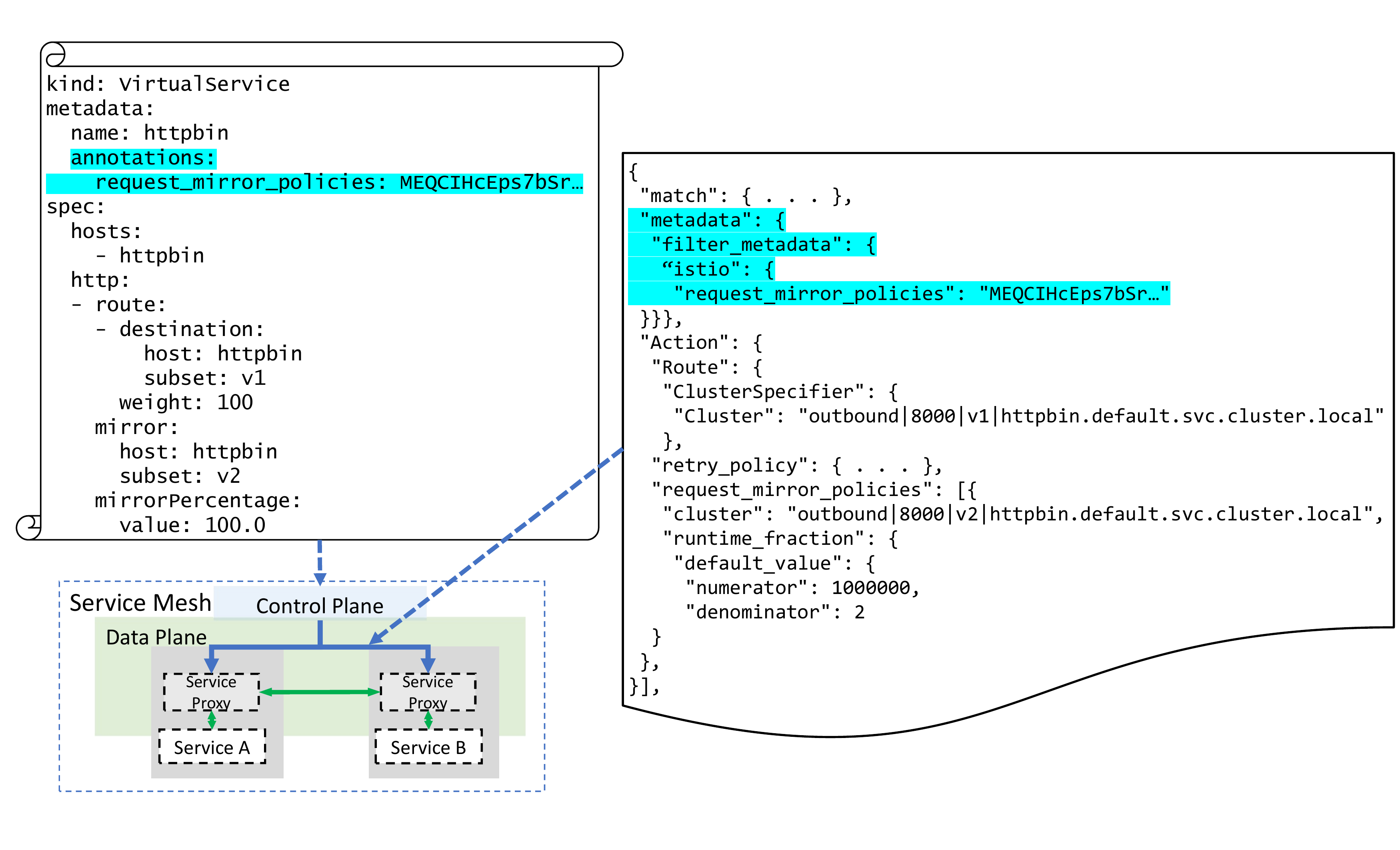}
\caption{Delivering the Signatures to the Envoy Proxies}
\label{fig:config-sign-transport}
\end{figure*}

The Application Owner defines the sidecar proxy configuration through CRDs.  A Configuration Signing Tool translates the configuration settings defined in the CRDs into the format used by the xDS API messages (where the xDS API is the interface between the control and the data plane of the service mesh).  The Configuration Signing Tool detects any policies in the Verifiable Configuration and signs the contents of those policies, as shown in figure~\ref{fig:config-sign-tool}.  The Application Owner adds to the CRD the signatures generated by the Configuration Signing Tool.  The Control Plane is enhanced to include the signatures contained in the CRDs that define the resources sent to the Proxies in the data plane, as shown in Figure~\ref{fig:config-sign-transport}.

\subsubsection{Verifying Configuration Received from the Control Plane}
\label{subsubsec:verify-config}
To verify the configurations received from the control plane, the service proxies contain a message filter located between the control plane and the proxy under the control of the service proxy/application image builder.  During secure bootup, the message filter receives the public key of the Application Owner and the Verifiable Configuration, which includes the list of policies that require signature verification.  The message filter unmarshals the incoming messages from the control plane during operation and gets a JSON representation of the resources. Next, it checks, using the Verifiable Configuration, if any parts of the resources need a signature verification. The resource configuration for which no signature verification is required is passed to the proxy. The resource configuration for which signature verification is required is forwarded to the proxy only if it has an associated signature, and the signature is verified using the public key of the Application Owner. If the signature verification fails for a resource configuration, that configuration is discarded.




\section{Discussion}
\label{sec:discussion}
\subsection{Security Applicability}
\label{subsec:security-applicability}

\begin{figure*}[h!]
\centering
\includegraphics[width=\linewidth]{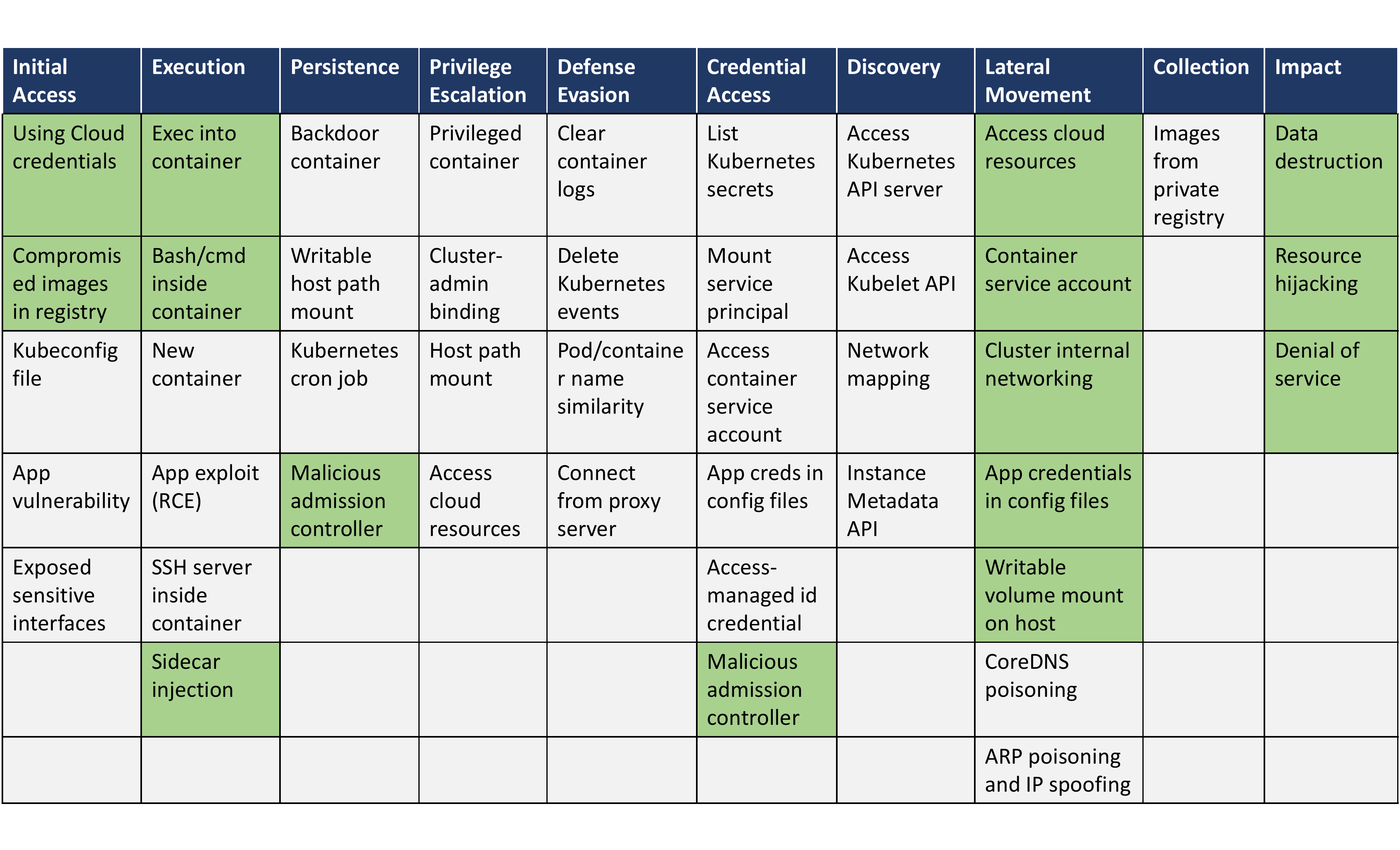}
\caption{Evaluation of our solution against the threat matrix for Kubernetes. The attack techniques highlighted in green are addressed in our solution}
\label{fig:threat-matrix}
\end{figure*}

The Verifiable Configuration allows the Application Owner to check and approve the settings of the resources that are sent to the XDS API: listeners, route configurations, scoped route configurations, virtual hosts, clusters, cluster load assignments, secrets, and runtime.  Control over these objects ensures that authorization, authentication, traffic management, and observability for the Service Mesh are done in a secure way that conforms with the policies defined by the Application Owner.  When used in conjunction with the Confidential Containers described in Section~\ref{sec:overview}, Verifiable Configuration enables and supports a wide array of security capabilities.   In this section, we use the Kubernetes threat matrix described in~\cite{kubernetes-threat-matrix}~and~\cite{kubernetes-threat-matrix-updated} to evaluate how our solution helps defend against the attack techniques defined in that document.

\subsection{Threat Matrix for Kubernetes}
\label{subsec:threat-matrix}
Microsoft Research initially released the threat matrix for Kubernetes in April 2020 and updated it a year later, in March 2021. It systematically maps the threat landscape of Kubernetes, following the structure of the MITRE ATT\&CK framework~\cite{mitre-attack}, the de-facto industry standard for describing threats. Each attack technique in this threat matrix is associated with one of these 10 MITRE ATT\&CK framework tactics: Initial access, Execution, Persistence, Privilege escalation, Defense evasion, Credential access, Discovery, Lateral movement, and Impact.

\subsection{Coverage of the Threat Matrix for Kubernetes}

Figure~\ref{fig:threat-matrix} provides an overview of the threat matrix for Kubernetes, highlighting the attack techniques against which Verifiable Configuration, used in conjunction with confidential containers, can defend an application running in the cluster.

Verifiable Configuration protects against two attack techniques from the Initial access tactic: Using Cloud credentials, and Compromised images in registry.  Attackers who have access to the cloud account credentials will behave like rogue Cluster Administrators.  Confidential containers are running inside security enclaves, as technologies like SEV, TDX, or IBM Z secure execution provide a Virtual Machine isolation boundary, making them inaccessible for an outsider.

In addition, Verifiable Configuration secures the xDS API interface, blocking all the ways of interacting with the data plane of the service mesh without the private key of the Application Owner, which is kept offline.  The Confidential Containers technology protects against compromised images in the registry because it provides a way to boot up encrypted images securely ~(\cite{Lum-encrypted-images},~\cite{confidential-containers}).  The Virtual Machine isolation boundary also protects applications against two Execution techniques: Exec into container and bash/cmd inside the container.  

Disabling automatic sidecar injection protects against the Sidecar injection technique.  Because we are securing the xDS API interface using the Verifiable Configuration, an external malicious admission controller running in the Kubernetes cluster will not change the state of the sidecar containers in the application service mesh.  Our solution is designed to protect the Application Owner from the Cluster Administrator by preventing an attacker who gained cluster admin rights from applying any Lateral movement attack techniques with respect to the application.  

Finally, Verifiable Configuration protects the applications against the three attack techniques in the Impact tactic. By addressing the scenario of CRD removal, it protects the application against Data destruction. Installing partial trust in the xDS API interface prevents any user other than the application owner from changing the resource configuration and therefore protects against Resource hijacking. Service Mesh provides defense against Denial of service attacks. Verifiable configuration protects the configuration of the sidecars, and prevents an adversary from tampering with it.

\section{Conclusion}
\label{sec:conclusion}
In this paper, we have defined the concepts of Verifiable Configuration and Partial Trust between the control plane and the data plane of a Service Mesh. We have shown that these two concepts allow the Application Owner to benefit from all the functions provided by the Service Mesh, while ensuring that the Cluster Administrator cannot change the state of the application in a way that the Application Owner did not intend. Using Istio, a popular Service Mesh implementation, as the motivating example, we have shown how Partial Trust can be created, established, identified, and enforced when creating, modifying, and deleting policies. Backed up by confidential containers and by the Cosign automated digital signing solution, Verifiable Configuration and Partial Trust defend the applications against multiple attack techniques identified in the Kubernetes Threat Matrix.

\bibliographystyle{unsrt}
\bibliography{references}

\end{document}